\documentclass[11pt]{article}

\newtheorem{lemma}{Lemma}
\newtheorem{proposition}{Proposition}
\newtheorem{theorem}{Theorem}
\newtheorem{corollary}{Corollary}

\def\tr{\mathop{\rm tr}\nolimits}

\def\ci{\mathop{\textrm{i}}\nolimits}

\begin{document}

\title{A Rainich-like approach to the Killing-Yano tensors} 
\author{Joan Josep Ferrando$^1$ and Juan Antonio S\'aez$^2$} 
\date{\today}          

\maketitle

\vspace{2cm}
\begin{abstract} 
The Rainich problem for Killing-Yano tensors posed by Collinson \cite{col} is solved. In intermediate steps, we first obtain the necessary and sufficient conditions for a 2+2 almost-product structure to determine the principal 2--planes of a skew-symmetric Killing-Yano tensor and then we give the additional conditions on a symmetric Killing tensor for it to be the square of a Killing-Yano tensor. We also analyze a similar problem for the conformal Killing-Yano and the conformal Killing tensors. Our results show that, in both cases, the principal 2--planes define a maxwellian structure. The associated Maxwell fields are obtained and we outline how this approach is of interest in studying the spacetimes that admit these kind of first integrals of the geodesic equation.
\end{abstract}

\vspace{1cm}

\vspace{2cm}
$^1$ Departament d'Astronomia i Astrof\'{\i}sica, Universitat
de Val\`encia, E-46100 Burjassot, Val\`encia, Spain.\\
E-mail: {\tt joan.ferrando@uv.es}

$^2$ Departament de Matem\`atica Econ\`omico-Empresarial, Universitat de
Val\`encia, E-46071 Val\`encia, Spain\\
E-mail: {\tt juan.a.saez@uv.es}

\newpage

\section{\large INTRODUCTION}

A second rank {\it Killing-Yano tensor} is a skew-symmetric tensor $A_{\alpha \beta}$ satisfying the equation
\begin{equation} \label{KYd}
\nabla_{(\alpha} A_{\beta) \mu} = 0
\end{equation}
It is known (see, for example, \cite{kra} and references therein) that the vector $v = A(t)$ is constant along an affinely parameterized geodesic with tangent vector $t$. Then, the scalar $v^2$ is a quadratic first integral of the geodesic equation and, consequently, defines a second rank {\it Killing tensor}, that is, a symmetric tensor $K_{\alpha \beta}$ solution to the equation
\begin{equation} \label{Kd}
\nabla_{(\alpha} K_{\beta \mu)} = 0
\end{equation}
This Killing tensor $K$ is in fact the square of $A$. 

Thus, if $A$ is a Killing-Yano tensor, then $K=A^2$ is a Killing tensor. But the converse is not true for a generic Killing tensor. Then, a question naturally arises: what conditions a Killing tensor $K$ must satisfy in order to be the square of a Killing-Yano tensor? This question was established by Collinson \cite{col} who also pointed out that it poses a problem analogous to that studied by Rainich for the Maxwell fields.

The energy tensor $T$ associated with an electromagnetic field $F$ solution of the source-free Maxwell equations ({\it Maxwell field}), $\nabla \cdot F = 0, \ dF=0$, is divergence-free, $\nabla \cdot T = 0$. Conversely, if $T$ is a conserved symmetric tensor, what additional conditions must it satisfy in order to be the energy tensor of a Maxwell field? This problem was posed and solved by Rainich \cite{rai} for regular fields obtaining, as a consequence, a fully geometric characterization of the non-null Einstein-Maxwell solutions. It is worth pointing out that the Rainich work \cite{rai} also includes other interesting results about the principal planes of a non-null Maxwell field. More precisely, Rainich theory for the regular electromagnetic field consists of the following elements: (i) to write the source-free Maxwell equations in terms of intrinsic variables, namely, the eigenvalues and the principal structure of the electromagnetic field, (ii) to give the necessary and sufficient conditions on a 2+2 structure in order to be the principal structure of a Maxwell field, (iii) to express Maxwell equations for the energetic variables, such, to obtain the algebraic conditions and the additional differential restrictions for a conserved symmetric tensor to be the energy tensor of a Maxwell field, and (iv) to write the latter conditions, via Einstein equations, for the Ricci tensor considered as a metric concomitant.

The main goal of the Rainich article is, of course, to reach point (iv) which leads to the so called 'already unified theory' \cite{mis}. Nevertheless, the interest in writing Maxwell equations in terms of energetic variables (point (iii)) was afterwards outlined by Witten \cite{wit}, althought the electromagnetic field was not, necessarily, the source of the gravitational field. On the other hand, the Rainich results for the principal planes (point (ii)) have shown their usefulness in several situations \cite{debmc} \cite{cff} \cite{fsD} \cite{cfs}. All these different aspects of a Rainich theory have been considered for a perfect fluid energy tensor: the local thermal equilibrium condition has been expressed in terms of energetic variables and a fully geometrical description of the thermodynamic perfect fluid solutions has been obtained \cite{cf4}; in this case point (ii) implies a characterization using the unitary velocity of the fluid, which has been accomplished for the holonomic \cite{cf2} and the barotropic \cite{cf3} perfect fluids.

As we have mentioned above, Collinson \cite{col} analyzed in his work aspect (iii) of the Rainich problem for the Killing-Yano tensors. He gave algebraic intrinsic conditions for a tensor to be the square of a skew-symmetric tensor. He also wrote the Killing-Yano equations in terms of the eigenvalues and eigenvectors (point (i)) and he discriminated between equations derived from the Killing tensor equations and those that the Killing-Yano condition adds. Here, we improve the Collinson results in two ways. Firstly, we undertake the aspect (ii) of the Rainich theory characterizing the Killing-Yano almost-product structures and, secondly, we give both, the algebraic and the additional differential conditions for a Killing tensor $K$ to be the square of a Killing-Yano tensor, as explicit equations on $K$ itself and the metric tensor $g$.

A {\it conformal Killing-Yano tensor} $A$ is a solution to the conformal invariant extension to the Killing-Yano equation (\ref{KYd}). Conformal Killing-Yano tensors define first integrals along affinely parameterized null geodesics with tangent vector $k$ and, in particular, $A^2(k,k)$ is a quadratic one. So, the square $P=A^2$ is a {\it conformal Killing tensor}. But there are conformal Killing tensors that are not the square of conformal Killing-Yano ones. Consequently, in this case we can state a Rainich-like problem similar to the one previously posed for the Killing-Yano tensors, a question that we also analyze here in the three aspects remarked above.

In studying both Rainich problems, for the Killing-Yano and for the conformal Killing-Yano tensors, we show that the underlying 2+2 structures are maxwellian and, in both cases, we obtain the associated Maxwell fields. This fact leads us to analyze the Rainich results about the electromagnetic field exhaustively, not only to better understand its different aspects, but also to introduce notation and concepts that enable us to fulfill the objectives of our work.

The spacetimes admitting second rank Killing-Yano tensors were considered by Collinson in his article showing that, in the vacuum case, the Weyl tensor is Petrov type D, N or O \cite{col}. This result was generalized later by Stephani for the non vacuum case \cite{ste}. The same restrictions for the existence of solutions to the conformal Killing-Yano equation have been shown more recently \cite{glkr}. The integrability conditions of the Killing-Yano equations and some of their consequences were analyzed by Dietz and R\"udiger \cite{diru2}, who also studied the canonical form of families of metrics admitting second rank Killing-Yano tensors \cite{diru3}. We can also quote the work by Hall \cite{hall} about Killing-Yano tensors in General Relativity.  The role played in these results by the maxwellian character of the 2+2 structure associated with the Killing-Yano tensor is pointed out in the present work. Some comments about the intrinsic characterization of these families of spacetimes are also presented.

In section 2 of the present paper we summarize in appropriate form the original Rainich theory for the non-null electromagnetic field in order to gain better understanding of its different aspects and, at once, we present notation and some essential results about 2+2 spacetime structures. In section 3 we solve the Rainich-like problem for the conformal Killing-Yano tensors. A similar Rainich problem for the Killing-Yano tensors is undertaken in section 4 improving, in this way, the Collinson results concerning this subject. Finally, section 5 is devoted to pointing out the interest of our results in characterizing the spacetimes that admit Killing-Yano or conformal Killing-Yano tensors.

\section{\large A SUMMARY OF THE RAINICH THEORY}

On an oriented spacetime $(V_4,g)$ of signature $(-+++)$ a 2+2 almost-product structure is defined by a time-like plane $V$ and its space-like orthogonal complement $H$. Let $v$ and $h= g-v$ be the respective projectors and let $\Pi$ be the {\it structure tensor}, $\Pi = v-h$. Almost-product structures can be classified by taking into account the invariant decomposition of the covariant derivative of $\Pi$ \cite{nav} or, equivalently, according to the foliation, minimal or umbilical character of each plane \cite{olga} \cite{fsD}. We will say that a structure is integrable when both planes are foliation and we will say that it is minimal or umbilical if both planes are so.

A 2+2 spacetime structure is also determined by the {\it canonical} unitary 2-form $U$, volume element of the time-like plane $V$. Then, the respective projectors are $v=U^2$ and $h = -(*U)^2$, where $U^2 = U \times U= \tr_{23} U \! \otimes \! U$ and $*$ is the Hodge dual operator. 

When both planes have a specific differential property, it is useful to introduce the self-dual unitary 2--form 
${\cal U} \equiv \frac{1}{\sqrt{2}} (U - {\rm {i}} *U )$ associated with $U$. The metric on the self-dual 2--forms space is ${\cal G} = \frac{1}{2} ( G - {\rm i} \eta )$, where $\eta$ is the metric volume element of the spacetime and $G$ is the metric on the space of 2--forms, $G=\frac{1}{2} g \wedge g$, $\wedge$ denoting the double-forms exterior product, $(A \wedge B)_{\alpha \beta \mu \nu} = A_{\alpha \mu} B_{\beta \nu} + A_{\beta \nu} B_{\alpha \mu} - A_{\alpha \nu} B_{\beta \mu} - A_{\beta \mu} B_{\alpha \nu}$. Then, we can consider some first order differential concomitants of ${\cal U}$ that determine the geometric properties of the structure \cite{fsD}. Indeed, if
$i(\cdot)$ denotes the interior product and $\delta$ the exterior codifferential, $\delta = *d*$, we have the following lemma \cite{fsD} 

\begin{lemma}  \label{lem-imu}
Let us consider the 2+2 structure defined by ${\cal U}=\frac{1}{\sqrt{2}} (U - i *U)$. It holds: \\ 
(i) The structure is minimal if, and only if,
\begin{equation}
\Phi = 2 \mbox{Re} [i(\delta {\cal U}){\cal U}] = \Phi[U] \equiv i(\delta U) U - i(\delta *U)*U= 0   \label{min}
\end{equation}
(ii) The structure is integrable if, and only if, 
\begin{equation}
\Psi = 2 \mbox{Im} [i(\delta {\cal U}){\cal U}] =\Psi[U] \equiv  - i(\delta U)*U -  i(\delta *U) U  = 0   \label{int}
\end{equation}
(iii) The structure is umbilical, if, and only if,
\begin{equation}
\Sigma[U] \equiv \nabla {\cal U} - i(\delta {\cal U}){\cal U} \otimes {\cal U} - i(\delta {\cal U}){\cal G}=0  \label{umb}
\end{equation}
\end{lemma} 
When the three conditions in lemma \ref{lem-imu} hold we have a product structure and $U$ satisfies $\nabla U=0$. It is worth pointing out that the first order differential properties of a 2+2 structure admit a kinematical interpretation \cite{cf1} and, in particular, the umbilical nature is equivalent to the geodesic and shear-free character of the two principal null directions of the structure \cite{fsD}.
 
On the other hand, taking into account expressions (\ref{min}) (\ref{int}) and considering the fact that $\Pi = U^2+*U^2$, a straightforward calculation allows us to write the first-order differential concomitants $\Phi$ and $\Psi$ in terms of the structure tensor $\Pi$:
\begin{eqnarray}
\Phi & = & \Phi[\Pi] \equiv - \frac{1}{2} \Pi(\nabla \cdot \Pi)  \label{PhiPi}  \\
\Psi & = & \Psi[\Pi] \equiv  \frac{1}{4} *(\nabla \Pi \times \Pi)  \label{PsiPi}
\end{eqnarray}
In equation (\ref{PsiPi}) and in the following, we put $*t$ to indicate the action of the Hodge dual operator on the skew-symmetric part of a tensor $t$.
Moreover the umbilical condition can also be expressed as a restriction on the structure tensor. Indeed, let us consider the totally symmetric tensor $\sigma=\sigma[\Pi]$:
\begin{equation}
\sigma[\Pi] \equiv {\cal S}\{2\nabla\Pi + \Pi(\nabla \cdot \Pi) \otimes \Pi - (\nabla \cdot \Pi) \otimes g \}            \label{sigma}
\end{equation}
where ${\cal S}\{t\}$ denotes the total symmetrization of a tensor $t$.
Then, we have:
\begin{lemma}  \label{lem-imu2}
Let $\Pi$ be a 2+2 structure tensor. It holds: \\ 
(i) The structure is minimal if, and only if, $\Phi[\Pi] = 0$\\
(ii) The structure is integrable if, and only if, $\Psi[\Pi] = 0$\\
(iii) The structure is umbilical, if, and only if, $\sigma[\Pi] = 0$
\end{lemma} 
The last property in lemma \ref{lem-imu2} can be directly inferred applying the geometric definition of umbilical structure \cite{fsD}. It also follows from the Dietz and R\"udiger results about the tensors with two geodesic and shear-free null principal directions \cite{diru1}. 

\subsection{Maxwell-Rainich equations}

A regular 2-form $F$ takes the canonical expression $F = e^{\phi}[\cos \psi U \! + \sin \psi \!*\!U]$, \newline where the {\it canonical unitary 2--form} $U$ determines the 2+2 associated structure, $\phi$ is the {\it energetic index} and $\psi$ is the {\it Rainich index}. 

Let us go on to the first point (i) of the Rainich work. We must express Maxwell equations in terms of the {\it canonical elements} $(U,\phi,\psi)$. Let $F$ be a {\it Maxwell field}, that is, a solution of the source-free Maxwell equations, $\delta F =0$, $\delta *F =0$. The self-dual $2$--form ${\cal F}= \frac{1}{\sqrt{2}} (F- \ci *F)$ writes ${\cal F} = e^{\phi + \mbox{\rm i} \psi} \ {\cal U}$. Then, taking into account that $2 \, {\cal U}^{\,2} = g$, Maxwell equations $\delta {\cal F}=0$, write  
\begin{equation}
\mbox{d} (\phi + \mbox{\rm i} \psi) = 2 i(\delta {\cal U}) {\cal U} 
\end{equation}
The real and imaginary parts of this equation lead to the {\it Maxwell-Rainich equations} \cite{rai} \cite{deb1}:
\begin{proposition}  \label{pro-maxrai}
In terms of the {\it canonical elements} $(U,\phi,\psi)$ of a non-null Maxwell field, the source-free Maxwell equations, $\delta F =0$, $\delta *F =0$ write:
\begin{eqnarray} 
\mbox{\rm d} \phi & = & \Phi[U] \equiv i(\delta U) U - i(\delta *U)*U  \label{max1a} \\[2mm]
\mbox{\rm d} \psi & = & \Psi[U] \equiv - i(\delta U)*U -  i(\delta *U) U  \label{max1b}
\end{eqnarray}
\end{proposition} 

\subsection{Maxwellian structures}

Given a 2+2 spacetime structure with canonical 2--form $U$, every pair of functions $(\phi,\psi)$ complete the canonical elements defining a regular 2--form $F \equiv (U,\phi,\psi)$. Nevertheless, a given $U$ is not always the canonical 2--form of a Maxwell field. When $F$ is a non-null solution of the source-free Maxwell equations one says that its underlying 2+2 structure is maxwellian. Then, we can ask the following question: Is it possible to express, solely in terms of $U$ and its derivatives, the necessary and sufficient conditions for $U$ to define a {\it maxwellian structure}? The answer to this question is affirmative and we can easily find these conditions starting from the Maxwell-Rainich equations (\ref{max1a})(\ref{max1b}). Indeed, applying the Poincar\'e lemma to these equations, the Rainich theorem \cite{rai} follows:
\begin{theorem}  \label{teo-rainich1}
A unitary 2-form $U$ defines a maxwellian structure if, and only if, it satisfies:
\begin{equation}\label{max2} 
\mbox{\rm d} \Phi[U]  = 0 ; \qquad  \qquad  \qquad
\mbox{\rm d} \Psi[U]  = 0 
\end{equation}
Given a solution $U$ to these equations, there exist two functions $(\phi,\psi)$ such that $\mbox{\rm d} \phi = \Phi[U]$, $\mbox{\rm d} \psi = \Psi[U]$. Then $F = e^{\phi}[\cos \psi U + \sin \psi *U]$ is a regular Maxwell field.
\end{theorem}

The functions $\phi$ and $\psi$ that theorem 1 associates to a maxwellian structure $U$ can be obtained  up to an additive constant. So, the associated Maxwell solution $F$ is determined up to a constant factor and a constant duality rotation. This theorem covers the second aspect (ii) of the Rainich work.

The Maxwellian character of a 2+2 structure can be simply expressed saying that the complex 1-form $i(\delta {\cal U}){\cal U}$ is closed:
\begin{equation}
\mbox{d} i(\delta {\cal U}) {\cal U} = 0
\end{equation}

\subsection{Maxwell equations for the energy tensor}

The energy ({\it Maxwell-Minkowski}) tensor $T$ associated to an electromagnetic field $F$ is minus the traceless part of its square and, for a regular field, it only depends on the canonical elements $(U,\phi)$ and can be expressed as:
\begin{equation}
T \equiv -\frac{1}{2}[F^2+*F^2] = -\frac{1}{2}e^{2\phi}[U^2+*U^2] 
= -\frac{1}{2}e^{2\phi}\Pi   \label{TF}
\end{equation} 
A simple calculation shows that the traceless tensor $T$ has a non-null square proportional to the metric:
\begin{equation}
\tr T = 0, \qquad \qquad   4 T^2 = \tr T^2\, g \not= 0 \label{alg1}
\end{equation}
Conversely, if a symmetric tensor satisfies the algebraic conditions (\ref{alg1}), we can obtain a simple 2--form $F^{\circ}$ as 
\begin{equation}
F^{\circ} = F^{\circ}(T) \equiv  \frac{Q(X)}{\sqrt{2Q(X,X)}} \, , \qquad  Q \equiv T \wedge g -\frac{1}{\sqrt{\tr T^2}} (T \wedge g)^2 \,   \label{alg2}  
\end{equation}
where $X$ is an arbitrary 2--form, and $R^2$ means the square of a double 2--form $R$ considered as an endomorphism on the 2--form space. Then, for an arbitrary Rainich index $\psi$, the 2--form $F = \cos \psi F^{\circ} + \sin \psi *F^{\circ}$ has $T$ as its energy tensor.

In order to guarantee the physical meaning of an energy tensor $T$ we must also impose the energy conditions on it. Under the algebraic restrictions (\ref{alg1}) the Pleba\'nski energy conditions reduce to:
\begin{equation}
T(x,x) > 0    \label{ec} 
\end{equation}
where $x$ is an arbitrary time-like vector. 

For a tensor given by (\ref{TF}) we have that $\nabla \cdot T = i(\delta F)F + i(\delta *F)F$ and so Maxwell equations imply that $T$ is divergence-free:
\begin{equation}
\nabla \cdot T = 0    \label{nablaT} 
\end{equation}
But the divergence-free condition (\ref{nablaT}) does not imply that any 2--form  having $T$ as its energy tensor is a Maxwell field. In order to undertake the point (iii) of the Rainich theory we must obtain the additional differential conditions on $T$ that complete its maxwellian character. We can write the conservation equation (\ref{nablaT}) in terms of the canonical energetic variables $(U,\phi)$. Indeed, from (\ref{TF}) and expression (\ref{PhiPi}) it follows that (\ref{nablaT}) is equivalent to the first Maxwell-Rainich equation (\ref{max1a}):
\begin{equation}
\mbox{\rm d} \phi = \Phi(\Pi)      \label{nablaT2} 
\end{equation}

It is worth pointing out that the conservation condition admits also a formulation in the sole structure tensor $\Pi$. If we name the 2+2 structure underlying to a conserved Maxwell-Minkowski energy tensor {\it pre-maxwellian structure} \cite{deb2}, it follows from (\ref{nablaT2}):

\begin{lemma}  \label{lemma-premax}
A structure tensor $\Pi$ defines a pre-maxwellian structure if, and only if, it satisfies:
\begin{equation}  \label{premax} 
\mbox{\rm d} \Phi(\Pi)  = 0 
\end{equation}
Given a solution $\Pi$ to this equation, there exists a function $\phi$ such that $\mbox{\rm d} \phi = \Phi(\Pi)$. Then $T = C e^{2\phi}\Pi$ is a conserved Maxwell-Minkowski energy tensor, $C$ being an arbitrary negative constant.
\end{lemma}

This last result is not a part of the Rainich work and we present it here for sake of completeness. Elsewhere \cite{cfs} we have studied a similar question for the Killing and the conformal Killing tensors.

Let us go on to the conditions that the whole set of Maxwell equations imposes on $T$. From proposition \ref{pro-maxrai} and expression (\ref{nablaT2}) for the conservation condition, besides this last equation we must impose the second Maxwell-Rainich equation (\ref{max1b}). In it we find the Rainich index $\psi$ that is not an energy variable. But we can eliminate it by differentiation and we obtain the second condition (\ref{max2}) of the Rainich theorem \ref{teo-rainich1}: $\mbox{\rm d}\Psi=0$. We know the expression (\ref{PsiPi}) of the 1--form $\Psi$ in terms of $\Pi$ and, taking into account (\ref{TF}), a straightforward calculation leads to 
\begin{equation}
\Psi = \Psi(T) \equiv \frac{1}{\tr T^2} *(\nabla T \times T)  \label{PsiT}
\end{equation}

With the results of this subsection we have acquired the point (iii) of the Rainich work that we present here as a second Rainich theorem \cite{rai} \cite{wit}:

\begin{theorem}  \label{teo-rainich2}
A symmetric tensor $T$ is the energy tensor of a Maxwell field if, and only if, it satisfies the algebraic conditions:
\begin{equation}
\tr T = 0, \qquad \quad   4 T^2 = \tr T^2\, g \not= 0, \qquad \quad T(x,x) > 0   \label{alg}
\end{equation}
and the differential ones:
\begin{equation}
\nabla \cdot T = 0, \qquad \quad   \mbox{\rm d} \Psi(T) =0    \label{dif}
\end{equation}
where the Rainich 1--form $\Psi(T)$ is given in (\ref{PsiT}) and $x$ is an arbitrary time-like vector.\\
Given a solution $T$ to these equations, there exists a function $\psi$ such that $\mbox{\rm d} \psi = \Psi(T)$. Then, if $F^{\circ}$ is given by (\ref{alg2}), $F = \cos \psi F^{\circ} + \sin \psi *F^{\circ}$ is a regular Maxwell field.
\end{theorem}

Let us note that the Rainich index $\psi$ is fixed up to an additive constant and, consequently, the Maxwell field associated to an energy tensor $T$ satisfying the conditions of theorem \ref{teo-rainich2} is determined up to a constant duality rotation.

In the following sections we analyze the three aspects of the Rainich theory for the Killing-Yano tensors and Conformal Killing-Yano tensors. However, it is worth pointing out that the Rainich work contains a last aspect which is its main goal: to give a fully geometric characterization of the non-null Einstein-Maxwell solutions. Nevertheless, this question easily follows on from theorem \ref{teo-rainich2}. Indeed, in dealing with Einstein-Maxwell solutions, $T$ coincides with the Ricci tensor because it is a traceless tensor. Moreover, the conservative condition for $T$ is a direct consequence of the Einstein equations. So, one must impose the Rainich algebraic conditions (\ref{alg}) and the Rainich equation $\mbox{\rm d} \Psi(Ric(g)) =0$ on the Ricci tensor (considered as a second order metric concomitant).

We will see that the 2+2 structures defined by a Killing-Yano tensor and a conformal Killing-Yano tensor are maxwellian. Moreover, every Killing-Yano tensor is a conformal one. So, in order to consider more and more restricted situations, we first analyze the conformal case and we finish with the Rainich theory for the Killing-Yano tensors.

\section{\large RAINICH THEORY FOR THE CONFORMAL KILLING-YANO TENSORS}

A Conformal Killing-Yano (CKY) tensor is a skew-symmetric tensor $A$ solution to the conformal invariant extension to the Killing-Yano equation (\ref{KYd}). This conformal Killing-Yano equation writes \cite{tach}:
\begin{equation} \label{CKYd}
\nabla_{(\alpha} A_{\beta) \mu} =  g_{\alpha \beta} a_{\mu} - a_{(\alpha} g_{\beta) \mu }
\end{equation}
where the 1--form $a$ is given by the codifferential of $A$: $3a = - \delta A$. 

If $A$ is a CKY tensor, the scalar $\kappa = A(k,p)$ is constant along an affinely parameterized null geodesic with tangent vector $k$, where $p$ is a vector orthogonal to  the geodesic and satisfying $k \wedge \nabla_k p =0$. In particular, we can take $p = A(k)$, which satisfies these restrictions as a consequence of the CKY equation. Then, the scalar $A^2(k,k)$ is a quadratic first integral of the null geodesic equation, so that, $P= A^2$ is a second rank Conformal Killing (CK) tensor, that is, a symmetric tensor solution to the CK equation
\begin{equation} \label{CKd}
\nabla_{(\alpha} P_{\beta \mu)} = g_{(\alpha \beta} b_{\mu)}
\end{equation}
It is worth mentioning that if $P$ is a CK tensor, so $P+ fg$ is for an arbitrary function $f$, and both define the same first integral of the null geodesic equation. So, we can always consider traceless CK tensors. In this case $b$ is in fact the divergence of $P$: $3b = \nabla \cdot P$.

\subsection{Conformal Killing-Yano equations in the variables $(U,\phi,\psi)$}

The first point (i) of the Rainich theory implies giving an expression of the CKY equation (\ref{CKYd}) in terms of the canonical elements $(U,\phi,\psi)$ of a regular CKY tensor $A = e^{\phi}[\cos \psi U + \sin \psi *U]$. In order to carry out this task it will be usefull to consider an alternative statement of the CKY condition (\ref{CKYd}) similar to those considered in \cite{diru2} for the KY equation. In fact, from the invariant decomposition of the covariant derivative $\nabla A$, it follows that (\ref{CKYd}) is equivalent to: 
\begin{equation} \label{CKYd2}
3 \nabla A =  dA - g \wedge \delta A
\end{equation}
where, for a vector, $v$ we put $(g \wedge v)_{\gamma \alpha \beta} = g_{\gamma \alpha} v_\beta - g_{\gamma \beta} v_{\alpha}$. From expression (\ref{CKYd2}) it follows that the CKY condition is invariant under Hodge duality, so that, $*A$ is a CKY tensor too. Consequently, the self-dual $2$--form ${\cal A}= \frac{1}{\sqrt{2}} (A- \ci *A)$ satisfies the CKY condition (\ref{CKYd2}) which in the self-dual formalism takes the form:
\begin{equation} \label{CKYd3}
3 \nabla {\cal A} =  2 i(\delta {\cal A}) {\cal G}
\end{equation}
Now we can put ${\cal A} = e^{\phi + \mbox{\rm i} \psi} \ {\cal U}$ in equation (\ref{CKYd3}). Then, we obtain an equation that can be partially decoupled. Indeed, its orthogonal to ${\cal U}$ part leads to a condition involving the sole variable $U$ which expresses precisely that the associated 2+2 structure is umbilical: $\Sigma[U]=0$. On the other hand, its component in the ${\cal U}$--direction leads to $i(\delta{\cal U}){\cal U} = - d(\phi + \mbox{\rm i} \psi)$. Thus, taking into account expressions (\ref{min}-\ref{umb}), we have shown:
\begin{proposition}  \label{pro-CKY}
In terms of the {\it canonical elements} $(U,\phi,\psi)$ of a non-null skew-symmetric tensor $A$, the CKY equation (\ref{CKYd}), write:
\begin{eqnarray}{ll}
\Sigma[U] \equiv \nabla {\cal U} - i(\delta {\cal U}){\cal U} \otimes {\cal U} - i(\delta {\cal U}){\cal G}=0      \label{CKY1a} 
\\[2mm]
-2 \mbox{\rm d} \phi  =  \Phi[U] \equiv i(\delta U) U - i(\delta *U)*U & \label{CKY1b} \\[2mm]
-2 \mbox{\rm d} \psi  =  \Psi[U] \equiv - i(\delta U)*U -  i(\delta *U) U & \label{CKY1c}
\end{eqnarray}
\end{proposition} 

\subsection{Conformal Killing-Yano structures}

It is evident that the CKY equations (\ref{CKY1a}-\ref{CKY1c}) admit an equivalent formulation in terms of the sole variable $U$ and, consequently, we can characterize the 2+2 structures associated with a CKY tensor. Indeed, if we name them {\it conformal Killing Yano structures}, a result similar to the first Rainich theorem follows from proposition \ref{pro-CKY}:
\begin{theorem}  \label{teo-CKY1}
The 2+2 CKY structures are the umbilical and maxwellian structures. That is, a unitary 2-form $U$ defines a CKY structure if, and only if, it satisfies:
\begin{equation} \label{CKY2}
\Sigma[U]=0 \, ;    \qquad  \qquad
\mbox{\rm d} \Phi[U]  = 0 \, , \qquad  \qquad  
\mbox{\rm d} \Psi[U]  = 0 
\end{equation}
Given a solution $U$ to these equations, there exist two functions $(\phi,\psi)$ such that $-2 \mbox{\rm d} \phi = \Phi[U]$, $-2 \mbox{\rm d} \psi = \Psi[U]$. Then $A = e^{\phi}[\cos \psi U + \sin \psi *U]$ is a regular CKY tensor.
\end{theorem}

This theorem covers the second aspect (ii) of the Rainich theory. The CKY tensors $A$ asociated to a CKY structure $U$ solution to the equations (\ref{CKY2}) are determined up to a constant factor and a constant duality rotation.

The maxwellian character of a CKY structure offers another interpretation for the CKY tensors: they are associated with a class of Maxwell fields, those having an umbilical underlying structure. More precisely, we have:
\begin{corollary}
A skew-symmetric tensor $A = e^{\phi}[\cos \psi U + \sin \psi *U]$ is a CKY tensor if, and only if, $F = e^{-2\phi}[\cos 2\psi U - \sin 2\psi *U]$ is an umbilical Maxwell field.
\end{corollary} 

\subsection{Conformal Killing-Yano equations for its traceless square}

The traceless square $P$ of a CKY tensor $A$ can write in terms of the canonical elements $(U,\phi)$:
\begin{equation}
P \equiv A^2 - \frac{1}{4} \tr A^2 g = \frac{1}{2}e^{2\phi}[U^2+*U^2] 
= \frac{1}{2}e^{2\phi}\Pi   \label{PA}
\end{equation} 
The algebraic characterization of $P$ is given by the algebraic Rainich conditions (\ref{alg1}) together with a condition imposing that the time-like eigenvalue is positive, $P(x,x) < 0$, $x$ being an arbitrary time-like vector.
 
If $A$ is a solution to the CKY equation (\ref{CKYd}), then $P$ is a CK tensor. But the CK condition (\ref{CKd}) does not imply that some CKY tensor $A$ has $P$ as its traceless square. To undertake point (iii) of the Rainich theory we must obtain the additional differential conditions on $P$ that complete its CKY character. In order to obtain these conditions, we start by writing the CK equation (\ref{CKd}) in terms of the variables $(\Pi,\phi)$. Putting the last expression of (\ref{PA}) in the CK equation we arrive to the conditions \cite{cfs}:
\begin{equation}
\sigma[\Pi] = 0 \, , \qquad \qquad  -2 \mbox{\rm d} \phi = \Phi(\Pi)      \label{CK2} 
\end{equation}
Taking into account lemmas \ref{lem-imu} and \ref{lem-imu2} and expression (\ref{PhiPi}) we find that the CK equations (\ref{CK2}) are equivalent to the two first equations (\ref{CKY1a}), (\ref{CKY1b}) of the CKY characterization given in proposition \ref{pro-CKY}.

The formulation (\ref{CK2}) for the CK conditions allows us to characterize the {\it conformal Killing structures}, that is, the structures associated with a CK tensor. This question has been analyzed elsewhere \cite{cfs}, and here we present some results for completeness:

\begin{lemma}  \label{lemma-CK}
The conformal Killing structures are the pre-maxwellian and umbilical structures. That is, a structure tensor $\Pi$ defines a conformal Killing structure if, and only if, it satisfies:
\begin{equation}  \label{CK3} 
\sigma[\Pi] = 0 \, , \qquad \qquad  \mbox{\rm d} \Phi(\Pi)  = 0 
\end{equation}
Given a solution $\Pi$ to these equations, there exists a function $\phi$ such that $\mbox{\rm d} \phi = \Phi(\Pi)$. Then $P = C e^{2\phi}\Pi$ is a CK tensor, $C$ being an arbitrary constant.
\end{lemma}

Let us go on the conditions that all CKY equations impose on $P$. From proposition \ref{pro-CKY} and expression (\ref{CK2}) for the CK condition, we must also impose the equation  (\ref{CKY1c}). We can easily eliminate $\psi$ in this equation and, if we write the 1--form $\Psi$ in terms of $P$, we obtain a result that corresponds to the second Rainich theorem: 

\begin{theorem}  \label{teo-CKY2}
A symmetric tensor $P$ is the traceless square of a CKY tensor if, and only if, it satisfies the algebraic conditions:
\begin{equation}
\tr P = 0, \qquad \quad   4 P^2 = \tr P^2 g \not= 0, \qquad \quad P(x,x) < 0   \label{CKY-alg}
\end{equation}
and the differential ones:
\begin{equation}
{\cal S}\{3 \nabla P - g \otimes \nabla \cdot P\}=0  \, , \qquad \quad   \mbox{\rm d} \Psi(P) =0    \label{CKY-dif}
\end{equation}
where the Rainich 1--form $\Psi(P)$ is given in (\ref{PsiT}) and $x$ is an arbitrary time-like vector.\\
Given a solution $P$ to these equations, there exists a function $\psi$ such that $-2 \mbox{\rm d} \psi = \Psi(P)$. Then, if $A^{\circ} = F^{\circ}[-P]$ where $F^{\circ}[T]$ is given by (\ref{alg2}), $A = \cos \psi A^{\circ} + \sin \psi *A^{\circ}$ is a CKY tensor.
\end{theorem}

Let us note that the Rainich index $\psi$ is fixed up to an additive constant and, consequently, the CKY tensors associated to a symmetric tensor $P$ satisfying the conditions of theorem \ref{teo-CKY2} is determined up to a constant duality rotation. As a corollary, a symmetric tensor is the square of a CKY tensor if, and only if, its traceless part satisfies the conditions of the theorem above.

\section{\large RAINICH THEORY FOR THE KILLING-YANO TENSORS}

A Killing-Yano (KY) tensor is a skew-symmetric tensor $A$ solution to the Killing-Yano equation (\ref{KYd}). We have commented in the introduction about the first integrals defined by $A$ and its square $K=A^2$, which is a Killing tensor solution to the generalized Killing equation (\ref{Kd}).

It is known that if $K$ is a Killing tensor, its traceless part $P= K - \frac{1}{4} \tr K g$ is a CK tensor. Moreover, every KY tensor $A$ is a CKY tensor, and the KY equation also implies that $A$ is a co-closed 2--form. It is easily to show that these two conditions are sufficient too. Thus, the results in the beginning of section 3, allows us to state: $A$ is a KY tensor if, and only if, it satisfies:
\begin{equation} \label{KYd2}
3 \nabla {\cal A} =  2 i(\delta {\cal A}) {\cal G} \, , \qquad \qquad \delta A =0
\end{equation}
where ${\cal A}= \frac{1}{\sqrt{2}} (A- \ci *A)$ is the self-dual 2--form associated with $A$.

\subsection{Killing-Yano equations in the variables $(U,\phi,\psi)$}

Let us go on to express the KY equation (\ref{KYd2}) in terms of the canonical elements $(U,\phi,\psi)$ of a regular KY tensor $A = e^{\phi}[\cos \psi U + \sin \psi *U]$. We know that the first equation in (\ref{KYd2}) states that $A$ is a CKY tensor and it has been written in terms of the canonical elements in proposition \ref{pro-CKY}. So, we must only add the second equation $\delta A = 0$ by putting $A$ in terms of the canonical elements. Then we arrive to an equation involving first derivatives of the three elements $(U,\phi,\psi)$. But equations (\ref{CKY1b}) and (\ref{CKY1c}) in proposition \ref{pro-CKY} give, respectively, the derivatives of the energetic index $\phi$ and the Rainich index $\psi$ in terms of derivatives of $U$. So, we can finally obtain a condition that is algebraic on the scalars $(\phi,\psi)$ and differential on $U$. This condition together with those in proposition \ref{pro-CKY} are equivalent to the KY equation. Thus, we have acquired the first point of the Rainich theory:
\begin{proposition}  \label{pro-KY}
In terms of the {\it canonical elements} $(U,\phi,\psi)$ of a non-null skew-symmetric tensor $A$, the KY equation (\ref{KYd}), write:
\begin{eqnarray}
\Sigma[U] \equiv \nabla {\cal U} - i(\delta {\cal U}){\cal U} \otimes {\cal U} - i(\delta {\cal U}){\cal G}=0      \label{KY1a} 
\\[2mm]
-2 \mbox{\rm d} \phi = \Phi[U] \equiv i(\delta U) U - i(\delta *U)*U  \label{KY1b} \\[2mm]
-2 \mbox{\rm d} \psi = \Psi[U] \equiv - i(\delta U)*U -  i(\delta *U) U  \label{KY1c} \\[2mm]
\cos \psi \, \delta U + \sin \psi \; \delta*U = 0   \label{KY1d}
\end{eqnarray}
\end{proposition} 

\subsection{Killing-Yano structures}

Now we look for the equations characterizing a {\it Killing-Yano structure}, that is, the conditions  in the sole variable $U$ equivalent to the whole Killing-Yano equations (\ref{KY1a}-\ref{KY1d}). The first one is already a condition on $U$ and the second one is equivalent to $\mbox{d} \Phi[U] = 0$. In this case, it is not suitable to eliminate $\psi$ in (\ref{KY1c}) because the Rainich index appears in equation (\ref{KY1d}) too. Nevertheless, from this last equation we can calculate $\psi$ in terms of $U$ and we can impose (\ref{KY1c}) on it. All these considerations lead to the following theorem:
\begin{theorem}  \label{teo-KY1}
A unitary 2-form $U$ defines a non-product Killing-Yano structure if, and only if, it satisfies:
\begin{eqnarray} 
\Sigma[U]=0 \, ,    \qquad  \qquad
\mbox{\rm d} \Phi[U]  = 0 \, , \qquad  \qquad 
\delta U \wedge \delta *U = 0   \label{KY2a}          \\[2mm] 
2 \mbox{\rm d} \psi[U] + \Psi[U] = 0 \, , \qquad  \quad \psi[U] \equiv - \arctan \left\{\frac{i(x)\delta U}{i(x)\delta *U}\right\}     \label{KY2b}
\end{eqnarray}
where $x$ is an arbitrary vector such that $i(x)\delta U \not= 0$ or $i(x)\delta *U \not=0$.
Given a solution $U$ to these equations, there exists a function $\phi$ such that $-2 \mbox{\rm d} \phi = \Phi[U]$. Then $A = e^{\phi}\{\cos \psi[U]\, U + \sin \psi[U] *U\}$ is a regular KY tensor.
\end{theorem}

The KY tensors $A$ associated to a KY structure $U$ solution to the equations (\ref{KY2a}-\ref{KY2b}) are determined up to a constant factor. 

The non-product character of the structure in theorem \ref{teo-KY1} is a sufficient condition for the non simulataneous nullity of 1--forms $\delta U$ and $\delta *U$. Then, one can determine the Rainich index by (\ref{KY2b}). On the other hand, when $U$ defines a product structure, $U$ itself and $*U$ are two independent KY tensors. This property has been known for years \cite{hall} and it follows easily from proposition \ref{pro-KY}. Thus, in order to complete the second element of the Rainich theory for the KY tensors we must state the following:

\begin{proposition}  \label{pro-KY2}
Every product structure, $\nabla U = 0$, is a Killing-Yano structure. Then, the canonical form $U$ and its dual $*U$ are two independent Killing-Yano tensors.
\end{proposition}
  
It is evident that a KY structure is maxwellian and, consequently, a Maxwell field $F$ may be associated with a KY tensor $A$. In this case, besides the umbilical nature of the structure, the condition (\ref{KY1d}) must be imposed. So, we have:
\begin{corollary}
A skew-symmetric tensor $A = e^{\phi}[\cos \psi U + \sin \psi *U]$ is a KY tensor if, and only if, $F = e^{-2\phi}[\cos 2\psi U - \sin 2\psi *U]$ is an umbilical Maxwell field satisfying $\cos \psi \delta U + \sin \psi \delta*U = 0$.
\end{corollary} 

\subsection{Killing-Yano equations for its square}

The square $K$ of a KY tensor $A$ can write in terms of the canonical elements $(U,\phi,\psi)$:
\begin{equation}
K \equiv A^2 = e^{2\phi}[\cos^2 \psi \, U^2 + \sin^2 \psi \, *\!U^2] 
= \frac{1}{2}e^{2\phi}[\Pi + \cos 2\psi g]   \label{KA}
\end{equation} 
The intrinsic algebraic characterization of a symmetric tensor $K$ of the form (\ref{KA}) was given by Collinson \cite{col}. Now we easily put these conditions in an explicit form, that is, in terms of the metric and the tensor $K$ itself. Indeed, the traceless part $\, P = K - \frac{1}{4} \tr K\, g \,$ of $K$ must satisfy the Rainich conditions (\ref{alg1}), and the time-like eigenvalue must be positive, that is, $K(x,x) < 0$, $x$ being an arbitrary time-like vector.

Conversely, if a symmetric tensor satisfies these algebraic conditions, we can obtain a Rainich index $\psi$ as 
\begin{equation}
\psi = \psi(K) \equiv \arctan \left\{\frac{\sqrt{4b-a^2}-a}{\sqrt{2(2b-a^2)}}\right\} \,  \quad \quad  a \equiv \tr K \, , \quad b \equiv \tr K^2            \label{alg3} 
\end{equation}
and a simple 2--form $A^{\circ} = F^{\circ}[-P]$ where $F^{\circ}[T]$ is given by (\ref{alg2}) and $\, P = K - \frac{1}{4} \tr K\, g \,$. Then $A = \cos \psi A^{\circ} \pm \sin \psi *A^{\circ}$ are such that $A^2 = K$.
 
If $A$ is a solution to the KY equation (\ref{KYd}), then $K=A^2$ is a Killing tensor. But, as Collinson points out in his work \cite{col}, there exist Killing tensors which are not a KY square. Collinson asked: what additional differential conditions $K$ must satisfy to complete its KY character? In order to obtain these conditions, we can start by writing the Killing equation (\ref{Kd}) in terms of the variables $(U,\phi, \psi)$. Putting the expression (\ref{KA}) in (\ref{Kd}) we obtain the equations \cite{cfs}:
\begin{eqnarray}
\Sigma[U] = 0 \, , \qquad \qquad  -2 \mbox{\rm d} \phi = \Phi(U)       \label{K2a} \\[2mm]
- \sin 2 \psi \mbox{\rm d} \psi = \sin^2 \psi \, i(\delta \! *\!U) *\!U + \cos^2 \psi \, i(\delta U) U       \label{K2b}
\end{eqnarray}
Then, we find that the two first Killing equations (\ref{K2a}) are the two first equations (\ref{KY1a}), (\ref{KY1b}) of the KY characterization given in proposition \ref{pro-KY}, and the  third condition (\ref{K2b}) is a consequence of (\ref{KY1c}) and (\ref{KY1d}). Moreover, a straightforward calculation shows that (\ref{K2b}) and one of the expressions (\ref{KY1c}), (\ref{KY1d}) imply the other one. Consequently, in order to impose the whole KY condition we must add one of the equations (\ref{KY1c}), (\ref{KY1d}) to the Killing equation.

Until now we have discriminated between the restrictions that the Killing tensor equation imposes on the canonical elements and those that the KY condition adds. In his work, Collinson obtains a similar result by using another formalism \cite{col}. But in order to acquire the point (iii) of the Rainich theory in a similar way to the one we have presented above for Maxwell fields and CKY tensors, we must write the additional conditions that complete the KY character as explicit equations for $K$. 

However, previously we present the characterization of the {\it Killing structures}, that is, the Killing tensor associated structures. This result has been acquired elsewhere \cite{cfs}:

\begin{lemma}  \label{lemma-K}
A structure tensor $U$ defines a Killing structure if, and only if, it satisfies:
\begin{equation}  \label{CK3} 
\Sigma[U] = 0 \, , \quad \quad  \mbox{\rm d} \Phi(U)  = 0 \, , \quad \quad \mbox{\rm d} i(\delta U) U =  i(\delta U)U \wedge i(\delta *U)*U
\end{equation}
Given a solution $U$ to these equations, there exist two functions $\phi$ and $\psi$ such that $-2 \mbox{\rm d} \phi = \Phi(\Pi)$ and $\, \mbox{\rm d}\{e^{2 \phi} \sin^2 \psi\} = -e^{2 \phi} i(\delta U)U$. Then $K = C e^{2\phi}[\cos^2 \psi \, U^2 + \sin^2 \psi \, *\!U^2] + D g$ is a Killing tensor, $C$ and $D$ being arbitrary constants.
\end{lemma}

Let us go on the explicit conditions that the whole KY equations impose on $K$. From proposition \ref{pro-KY} and taking into account the comment after expression (\ref{K2b}), we can complete the KY condition by adding equation (\ref{KY1c}) to the Killing condition. We must give explicit expressions for $\psi$ and $\Psi$ in terms of $K$. The first one is the algebraic scalar invariant of $K$ given in (\ref{alg3}), and equation (\ref{PsiT}) gives the second one in terms of the traceless part of $K$. So, we finally arrive to the following theorem: 

\begin{theorem}  \label{teo-KY2}
A symmetric tensor $K$ is the square of a KY tensor if, and only if, it satisfies the algebraic conditions:  
\begin{equation}
4 P^2 = \tr P^2 g \not= 0 \, , \qquad \quad  P = K - \frac{1}{4} \tr K\, g \, ,    \qquad \quad   K(x,x) < 0   \label{KY-alg}
\end{equation}
and the differential ones:
\begin{equation}
{\cal S}\{ \nabla K \}=0  \, , \quad \quad   -2 \mbox{\rm d} \psi[K] = \epsilon \Psi[P]  \quad  \mbox{\rm for} \quad  \epsilon = 1 \quad \mbox{\rm or} \quad \epsilon = -1    \label{KY-dif}
\end{equation}
where the Rainich 1--form $\Psi(P)$ and the Rainich index $\psi[K]$ are given by (\ref{PsiT}) and (\ref{alg3}) respectively, and where $x$ is an arbitrary time-like vector.\\
Let $K$ be a solution to these equations. Then, if $A^{\circ} = F^{\circ}[-P]$ where $F^{\circ}[T]$ is given by (\ref{alg2}), $A = \cos \psi[K] A^{\circ} + \epsilon \sin \psi[K] *A^{\circ}$ is a KY tensor.
\end{theorem}

Let us note that if $K$ satisfies the algebraic conditions (\ref{KY-alg}) there exist two skew-symmetric tensors whose square is $K$, but only one of them can be a solution of the KY equations. 

\section{\large SPACETIMES ADMITTING KILLING-YANO OR CONFORMAL KILLING-YANO TENSORS}

It is known that a spacetime admitting a regular Killing-Yano tensor is, necessarily, type D or O and, in the first case, the Killing-Yano structure is aligned with the principal structure of the Weyl tensor \cite{ste}. A similar result is known for the CKY tensors \cite{glkr}. These properties can easily be obtained from our results about KY and CKY structures if, moreover, we take into account the integrability conditions for the umbilical character of a spacetime 2+2 structure. These conditions were considered using spinorial formalism by Dietz and R\"udiger \cite{diru1} in studying spacetimes admitting two geodesic and shear-free null congruences, and they have recently been revisited in tensorial formalism \cite{cfs}. In this last work we can find the following result:
\begin{lemma}
If a non conformally flat spacetime admits an umbilical and maxwellian 2+2 structure, then the Weyl tensor is type D and the structure is aligned with the Weyl principal structure.
\end{lemma} 
From this lemma and theorem \ref{teo-CKY1} we have:
\begin{corollary}  \label{cor-KYet}
If a non conformally flat spacetime admits a CKY tensor, then the Weyl tensor is type D and the principal structure is aligned with the CKY structure.
\end{corollary}
In particular, a KY tensor is a CKY tensor. So we recover the known result quoted above \cite{ste} \cite{glkr}.

Thus, our analysis to the underlying structures to the KY and CKY tensors is useful in studying the spacetimes where these 'symmetries' exist. Moreover, we can obtain not only the necessary condition given in corollary \ref{cor-KYet}, but we can also look for sufficient conditions obtaining, in this way, an intrinsic characterization of these families of spacetimes. Indeed, theorems \ref{teo-CKY1} and \ref{teo-KY1} allow us to state something more than corollary \ref{cor-KYet}:

\begin{proposition}  \label{pro-KYet}
A type D spacetime admits a CKY tensor if, and only if, its principal structure is umbilical and maxwellian, that is, its principal 2--form $U$ satisfies equations (\ref{CKY2}).\\[2mm]
A type D spacetime admits a KY tensor if, and only if, its principal 2--form $U$ satisfies equations (\ref{KY2a}) and (\ref{KY2b}).
\end{proposition}

It is worth pointing out that only the principal 2--form $U$ appears in the intrinsic characterization given in this proposition and an explicit expression for the metric concomitant $U$ is known \cite{fms}. So, we can write {\it intrinsic and explicit} conditions which can be tested by simple substitution of the metric tensor $g$ in order to know whether the spacetime admits a KY or a CKY tensor. Elsewhere \cite{fsS} we have commented on the interest in obtaining an intrinsic and explicit identification of a family of metrics. Moreover, given a metric $g$ verifying these equations, theorems \ref{teo-CKY1} and \ref{teo-KY1} say how the KY or the CKY tensor can be determined. A more detailed analysis about these questions and other sequels of this work will be considered elsewhere. 

\section*{\large ACKNOWLEDGMENTS}
The authors would like thank B. Coll for some useful comments. This work has been supported by the Spanish Ministerio de Ciencia y Tecnolog\'{\i}a, project AYA2000-2045.

\end{document}